\documentstyle[prc,aps,epsf]{revtex}
\def\g{\gamma}
\def\a{\Gamma}

\begin{document}
\draft
\title{
 Search for Supernarrow Dibaryons in $p d$ Interactions}
\author
{L.V.~Fil'kov$^1$, V.L.~Kashevarov$^1$, E.S.~Konobeevski$^2$,
M.V.~Mordovskoy$^2$, S.I.~Potashev$^2$ and V.M.~Skorkin$^2$} 
\address{
$^1$Lebedev Physical Institute, Moscow, Russia \protect\\
$^2$Institute for Nuclear Research, Moscow, Russia \\ }
\date{\today}
\maketitle
\begin{abstract}
The reaction $pd\to pX$ at 305 MeV is studied with the aim to search for
supernarrow dibaryons, the decay of which into two nucleons is forbidden 
by the Pauli exclusion principle. The experiment was  carried out at the 
Moscow Meson Factory using the spectrometer TAMS, which detected the 
scattered proton and another charged particle (either $p$ or $d$) from
the decay of $X$. Narrow peaks in missing mass spectra
have been observed at 1905 and 1924 MeV. Comparison of the obtained
data with theoretical predictions leads to the conclusion that
the peak found at 1905 MeV most likely corresponds to
a supernarrow dibaryon with the isotopic spin equal to 1.
The possible origin of the peak at 1924 MeV is also discussed.
\end{abstract}

\section{Introduction \protect\\}
\label{sec:intro}

The possibility of the existence of multiquark states was predicted by QCD
inspired models \cite{jaf,muld}. These works initiated a lot of experimental
searches for six-quark states (dibaryons). Usually one looked for
dibaryons in the NN channel.
In the present work we will consider supernarrow dibaryons (SNDs),
the decay of which into two nucleons is forbidden by the Pauli exclusion
principle \cite{fil1,akh,fil2,ger,alek,alek1}.
In the NN channel such dibaryon states correspond to
even singlets and odd triplets at the isotopic spin $T=0$,
and odd singlets and even triplets at $T=1$.
These dibaryons with the mass \mbox{$M < 2m_{N}+m_{\pi}$}
($m_N (m_{\pi}$) is the nucleon (pion) mass) can decay
into two nucleons, mainly emitting a photon. This is a new class of dibaryons
with decay widths \mbox{$\le 1$keV}. The
contribution of such dibaryons to strong interaction processes of hadrons
is small. However, their contribution to electromagnetic processes
on light nuclei may exceed the cross section for the process under study
out of range of the dibaryon resonance by several orders of magnitude
\cite{fil2,alek,alek1}.
The experimental discovery of such states would have important consequences
for particle and nuclear physics.

In the frame of the MIT bag model, Mulders et al. \cite{muld} calculated
the masses of different dibaryons, in particular, NN-decoupled dibaryons.
They predicted dibaryons $D(T=0;J^P =0^{-},1^{-},2^{-};M=2.11$ GeV) \ and
$D(1;1^{-};2.2$ GeV) corresponding to the forbidden NN states $^{13}P_J$
and $^{31}P_1$.
However, the dibaryon masses obtained exceed the pion production threshold.
Therefore, these dibaryons can decay into the $\pi NN$ channel. The possibility
of the existence of dibaryons with masses $M<2m_N+m_\pi $ was discussed by
Kondratyuk et al. \cite{kon1} in the model of stretched rotating bags, taking
account of spin-orbital quark interactions. In the frame of the chiral
soliton model, Kopeliovich \cite{kop} predicted that the masses of
$D(T=1,J^P=1^+)$ and $D(0,2^+)$
dibaryons exceeded the two-nucleon mass by 60 and 90 MeV, respectively.
These values are lower than the pion production threshold.

Unfortunately, all results obtained  for the dibaryon masses are model
dependent. Therefore, only an experiment could answer the question about
the existence of SNDs and their masses.

In the work \cite{bilg} the existence of a dibaryon, called $d'$,
with quantum numbers $T=even$ and $J^P=0^-$ which forbid its decay into
two nucleons, and with the mass $M=2.06$ GeV,
and the decay width $\a_{\pi NN}=0.5$ MeV, has been
postulated to explain the observed resonance-like
behavior in the energy dependence of the pionic double charge exchange
on nuclei at an energy below the $\Delta$-resonance.
However, there is a more convential interpretation of these data.
It was shown \cite{nus} in the frame of the distorted-wave impulse
approximation that such a peak arises nuturally because of the pion
propogation in the sequential process, in which pion double charge exhange
occurs through two successive $\pi N$ charge exchange reactions on two
neuterons.

In the work \cite{khr} dibaryons with exotic quantum numbers
were searched for in the process $pp\to pp \g\g$. The experiment was
performed
with a proton beam from the JINR phasotron at the energy of 198 MeV.
Two photons were detected in the energy range between of 10 and 100 MeV.
Some structure was observed in the photon energy spectrum, which was
attributed to the exotic dibaryon with the mass $M\approx 1920$ MeV.
The statistical significance of the effect is about 8$\sigma$.
However, a big width (FWHM=$31.3\pm 5.0$ MeV) of the observed structure
was obtained, and an additional careful experimental study of this
reaction is needed in order to understand the nature of this structure.
It is necessary to measure the photon energy spectrum with a higher energy
resolution and a higher statistical precision. Moreover, measurements
at two different incident proton energies are needed to check whether
the structure observed is a SND.

On the other hand, an analysis \cite{cal} of the Uppsala proton-proton
bremsstrahlung data looking for the presence of a dibaryon in the mass range
from 1900 to 1960 MeV only gave the upper limits of 10 and 3 nb for
the dibaryon production cross section at proton beam energies of 200 and
310 MeV, respectively. This result disagrees with the data \cite{khr}.

Earlier we have carried out measurements of missing mass spectra in the
reaction $pd\to pX$ \cite{konob} using a double-arm spectrometer at the
Moscow Meson Factory.
The narrow peak in the missing mass spectrum at 1905 MeV
with a width equal to the experimental resolution of 7 MeV has been observed
in this experiment.

As will be shown below, the nucleons and the deuteron from the decay of
the dibaryons under consideration into $\g NN$ and $\g d$
have to be emitted in a narrow angular cone with respect to the
direction of motion of the dibaryon. The size of this cone depends also on
the quantum numbers of the dibaryon. So a detection of the scattered proton
in coincidence with the proton (or the deuteron) from dibaryon decay
at correlated angles gives a good possibility to separate the SNDs
from the background and to determine their quantum numbers.

In the present paper we give a more detailed study of the reactions
$pd\to p+pX_1$ and $pd\to p+dX_2$ (where $X_1$ and $X_2$ are undetected
particles in this experiment) with the aim to search for SNDs
at the Moscow Meson Factory using an improved facility.
We will consider the following dibaryons:
$D(T=0,J^P=0^+)$, $D(0,0^-)$, $D(1,1^+)$, and $D(1,1^-)$.

It is worth noting that the  state $(T=1, J^P=1^-)$ corresponds
to the states $^{31}P_1$ and $^{33}P_1$ in the NN channel.
The former is forbidden and
the latter is allowed for a two-nucleon state. In our work we will study
the dibaryon $D(1,1^-)$, a decay of which into two nucleons is forbidden
by the Pauli principle (i.e. $^{31}P_1$ state).

The contents of the paper are the following. In Sec.\ \ref{sec:2} the decay
widths of SNDs are given and the cross sections of these dibaryon
production in the processes $pd\to pD$ are calculated. A description of
the experimental setup is in Sec.\ \ref{sec:setup}.
In Sec.\ \ref{sec:result} the results of the measurements are
presented. The results of a Monte Carlo calculation and an analysis of the
experimental data obtained are presented in Sec.\ \ref{sec:analys}.
The main conclusions are given in Sec.\ \ref{sec:concl}.

\section{Cross sections of the supernarrow dibaryon production
in the reaction $\lowercase{pd}\to \lowercase{p}D$ \protect\\}
\label{sec:2}

In the process $pd\to pD$, SNDs can be produced
only if the  nucleons in the deuteron overlap sufficiently, such that a
six-quark state with deuteron quantum numbers can be formed. In this case, an
interaction of a photon or a meson with this state can change its
quantum numbers so that a metastable state is formed.
Therefore, the probability of the production of such dibaryons is
proportional to the probability $\eta$ of the six-quark state existing
in the deuteron.

The magnitude of $\eta$ can be estimated from the
deuteron form factor at large $Q^2$ (see, for example, \cite{bur}). However,
the values obtained depend strongly on the model of the form factor of
the six-quark state over a broad region of $Q^2$. Another way to estimate
this parameter is to use the discrepancy between the theoretical and
experimental values of the deuteron magnetic moment \cite{kim,kon2}.
This method is free from the restrictions quoted above and gives
$\eta\le 0.03$ \cite{kon2}. We assume here that $\eta$
and the probability of the full overlap of nucleons in the virtual singlet
state $^{31}S_0$ are equal to 0.01.

Since the energy of nucleons, produced in the decay of the dibaryons under
study with $M<2m_N +m_{\pi}$, is small, it may be expected that the main
contribution to a two-nucleon system should come from $^{13}S_1$ (deuteron)
and $^{31}S_0$ (virtual singlet) states. The results of calculations of the
decay widths of the dibaryons into $\g d$ and $\g NN$ on the
basis of such assumptions are listed in Table \ref{table1}.

As shown in \cite{alek1} the main contribution to the dibaryon
photoproduction
from the deuteron is given by the one-pion exchange diagram. Therefore, we
will describe the production of SNDs in the process $pd\to p D$ using
the one meson exchange diagram also.

The production of the isoscalar dibaryons $D(0,0^+)$ and $D(0,0^-)$
will be calculated with the help of the pole
diagram with $\omega $-meson exchange between the incident proton and
the deuteron. The vertices $D(0,0^+)\to\omega d$ and $D(0,0^-)\to\omega d$
can be written as
\begin{equation}
\Gamma_{D(0,0^+)\to\omega d}=\frac{f_1}{M}\sqrt{\eta}
\epsilon_{\mu\nu\lambda\sigma}G^{\mu\nu}\Phi^{\lambda\sigma}_1,
\end{equation}
\begin{equation}
\Gamma_{D(0,0^-)\to\omega d}=\frac{f_2}{M}\sqrt{\eta}
G_{\mu\nu}\Phi^{\mu\nu}_1,
\end{equation}
where $G_{\mu\nu}=v_{\mu}r_{\nu}-r_{\mu}v_{\nu}$,
$\Phi_{1\mu\nu}=w_{\mu}p_{\nu}-p_{\mu}w_{\nu}$. Here $v$ and $w$ are the
four-vectors of the deuteron and $\omega$ meson polarization, $r$ and $p$
are their four-momenta.

 As a result of the calculations we have
\begin{equation}
\frac{d\sigma_{pd\to pD(0^+)}}{d\Omega }=
\frac{f_1^2}{4\pi }F_0R_{(+)},
\end{equation}
\begin{equation}
\frac{d\sigma_{pd\to pD(0^-)}}{d\Omega }=
\frac 14\frac{f_2^2}{4\pi }F_0R_{(-)},
\end{equation}
where
\[
F_0=\frac{g_{\omega NN}^2}{4\pi }\frac{F}{(t-m_{\omega}^2)^2},
\]
\[
F=\frac 83\eta \frac{p_2^2}{M^2}\frac{1}
{m_dp_1\left[\left( m_d+E_1\right) p_2-p_1E_2\cos \theta \right]},
\]
\begin{eqnarray*}
R_{(-)}&=&-\frac14 \{2t[(s-m_d^2-m^2_N)(s+t-M^2-m^2_N)+tm_d^2] \\
&&+(M^2-m_d^2-t)^2(2m^2_N+t)\},
\end{eqnarray*}
\[
R_{(+)}=R_{(-)}+tm_d^2(2m^2_N+t),
\]
\[
 t=2(m^2_N-E_1E_2+p_1p_2\cos\theta),
\]
$E_1,p_1(E_2,p_2)$ are the energy and momentum of the initial (final) proton,
$\theta $ is the angle of the scattered proton emission, $m_d$ is the
deuteron mass, $g_{\omega NN}^2$ is the constant of the $\omega N$
interaction,
$f_{1,2}^2/4\pi$
are the coupling constants in the vertices $D(0,1^{+})+\omega \to
D(0,0^+)$ and $D(0,1^+)+\omega \to D(0,0^-)$, respectively,
and $D(0,1^+)$ is the six-quark state with quantum numbers of the deuteron.

The isovector dibaryons $D(1,1^+)$ and $D(1,1^-)$
can be formed mainly as a result of pion
exchange between the incident proton and the deuteron.
The vertices $D(1,1^+)\to \pi+d$ and $D(1,1^-)\to \pi+d$ are written as
\begin{equation}
\a_{D(1,1+)\to\pi d}=\frac{g_1}{M}\sqrt{\eta}\epsilon_{\mu\nu\lambda\sigma}
G^{\mu\nu}\Phi_2^{\lambda\sigma},
\end{equation}
\begin{equation}
\a_{D(1,1-)\to\pi d}=\frac{g_2}{M}\sqrt{\eta}
G_{\mu\nu}\Phi_3^{\mu\nu},
\end{equation}
where $\Phi_{2\mu\nu}=\varepsilon_{1\mu}q_{1\nu}-q_{1\mu}\varepsilon_{1\nu}$,
$\Phi_{3\mu\nu}=\varepsilon_{2\mu}q_{2\nu}-q_{2\mu}\varepsilon_{2\nu}$, and
$\varepsilon_{1(2)}$ and $q_{1(2)}$ are the four-vectors of polarization and
four-momenta of the $D(1,1^+)$ ($D(1,1^-)$), respectively.

The calculations yield the following expressions for the cross section of
these dibaryon production
\begin{equation}
\frac{d\sigma_{pd\to pD(1^+)}}{d\Omega }=\frac{g_1^2}{4\pi }F_1
[(M^2+m_d^2-t)^2-4m_d^2M^2],
\label{one}
\end{equation}
\begin{equation}
\frac{d\sigma_{pd\to pD(1^-)}}{d\Omega}=\frac14\frac{g_2^2}{4\pi}F_1
[(M^2+m_d^2-t)^2+2m_d^2M^2],
\label{two}
\end{equation}
where
\[
F_1=-\frac14\frac{g_{\pi NN}^2}{4\pi}\frac{tF}{(t-m_{\pi}^2)^2},
\]
and $g_{1,2}^2/(4\pi )$ are the coupling constants in the vertex
for the transition of the six-quark state $D(0,1^+)$ into the dibaryons under
consideration via the pion exchange.

\begin{figure}
\vspace*{3cm}
\epsfysize=12cm
\epsfxsize=15cm
\epsffile{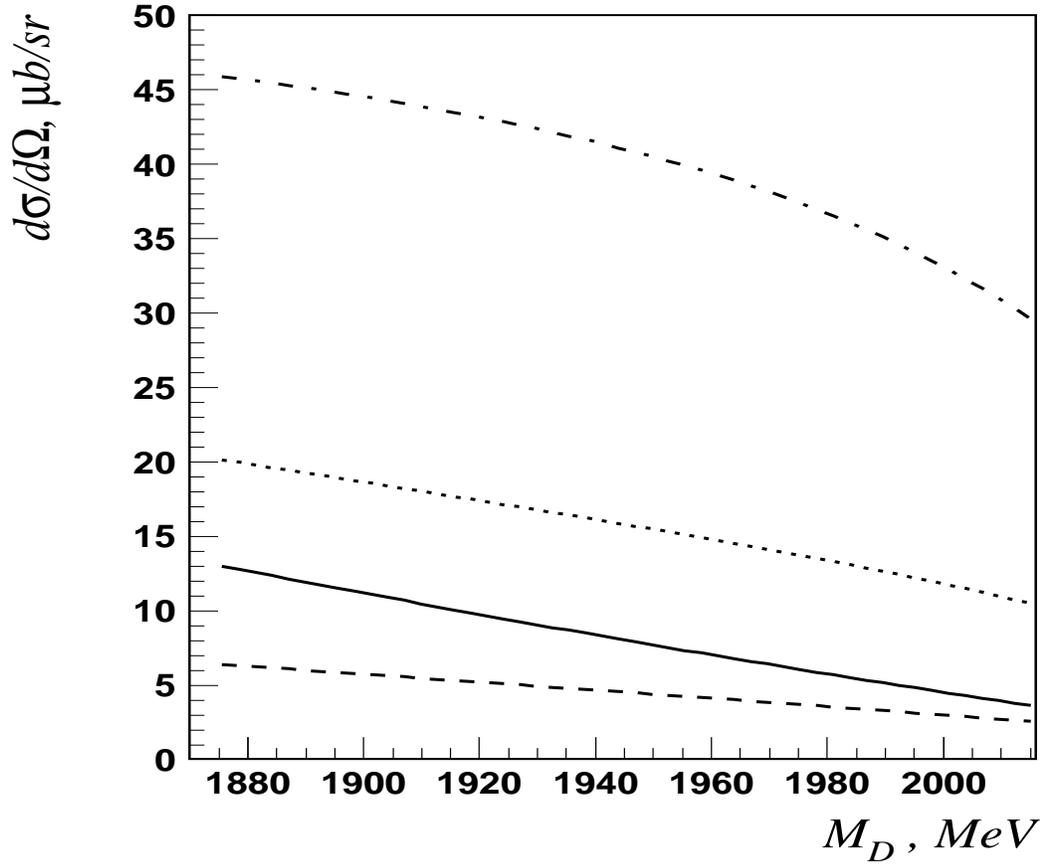}
\caption{ The cross section of the supernarrow dibaryon production
in the process $pd\to pD$ at $T_1=305$ MeV and $\theta=70^{\circ}$
as a function of the dibaryon mass $M$. The solid, dashed, doted and
dash-doted lines correspond to the production of the dibaryons
$D(0,0^+)$, $D(0,0^-)$, $D(1,1^+)$ and $D(1,1^-)$, respectively.
\label{fig1}}
\end{figure}

For a numerical calculation of the dibaryon contribution let us assume that
$g_{\pi NN}^2/4\pi =14.6$, $g_{\omega NN}^2/4\pi =19.2$, $\eta =0.01$.
The constants $g_{1,2}^2/4\pi $ and $f_{1,2}^2/4\pi $ are unknown.
They are strong interaction constants.
In order not to overestimate the
contributions of the dibaryons we put these constants equal to 1.
The results of the calculation  of the SND
production cross sections in the process $pd\to pD$
at the kinetic energy of the incident proton $T_1=E_1-m=305$ MeV and the
emission angle of the scattered proton $\theta =70^{\circ}$  as a
function of the dibaryon mass $M$ are shown in Fig.~\ref{fig1}. Here
the solid , dashed, dotted and dash-dotted lines  correspond to
$D(0,0^+)$, $D(0,0^-)$, $D(1,1^+)$, and $D(1,1^-)$ dibaryon production,
respectively.

As indicated above, the SNDs decay mainly emitting a
photon. Therefore, if we limit ourselves to the investigation of
$pd$ interactions with the photon in the final state, then the
contribution of such dibaryons will essentially exceed the cross sections
of the background processes.
On the other hand, the special choice of the $pd\to pX$ kinematics allows us
to allocate the area where the contribution of the SNDs
dominates, even without detection of the final photon.

\section{Experimental setup \protect\\}
\label{sec:setup}

The reaction $p+d\to p+X$ was studied at the proton accelerator of
the Moscow Meson Factory at 305 MeV.
A proton beam with an average effective
intensity $\sim 0.1$ nA bombarded alternatively CD$_2$
and $^{12}$C targets of 0.14 and 0.18 g/cm$^2$, respectively. The $pd$
reaction contribution was determined by subtracting the $^{12}$C
spectrum from the CD$_2$ spectrum. The exposition time of the experiment
using CD$_2$ as a target was 100 hours. This period consisted of two
runs corresponding to two different kinematic conditions in order to avoid a
systematic error. The charged particles produced in the reaction in
question were detected at different correlated angles by the Two Arms
Mass Spectrometer (TAMS).

\begin{figure}
\vspace*{1cm}
\epsfysize=20cm
\epsfxsize=18cm
\epsffile{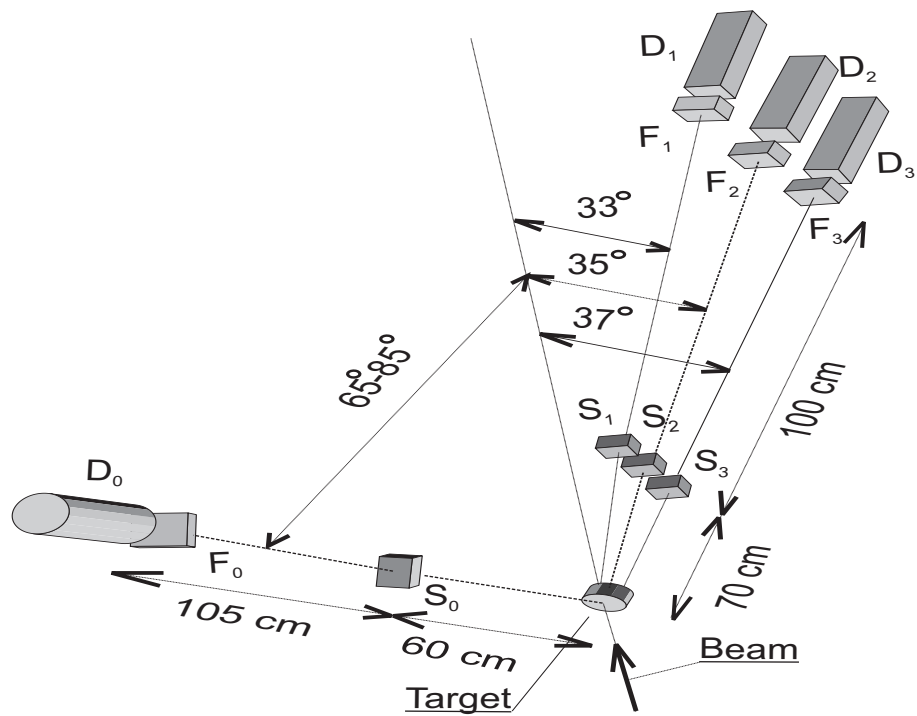}
\caption{The Two Arm Mass Spectrometer TAMS. $S_0$, $S_1$, $S_2$ and $S_3$
are start detectors; $F_0$, $F_1$, $F_2$ and $F_3$ are stop $\Delta E$
detectors; $D_0$ is a BGO detector; $D_1$, $D_2$ and $D_3$
are full absorption $E$ detectors.
\label{fig2}}
\end{figure}

Our layout is shown schematically in Fig.~\ref{fig2}. The left movable
spectrometer arm, which is a single telescope $\Delta E-\Delta E-E$, was used
to measure the energy and the time of flight of the scattered proton
at a fixed emission angle $\theta_L\equiv\theta$.
A remote control drive allowed to place this arm at various angles
between $65^{\circ}$ and $85^{\circ}$. In the present experiment
TAMS detected the scattered proton at the angle
$\theta_L= 72.5^{\circ}$ (and $70^{\circ}$ in another run)
in coincidence with the second charged particle
(either $p$ or $d$) from the decay of the particle $X$
during 55 (and 45) hours.

The detection of the second  charged particle  in the right arm at
angles close to the emission angle of particle $X$ with mass $M$
allows to suppress essentially the contribution of background processes
and increase the relative contribution
of a possible SND production. The right arm consisted of three telescopes
which were located at $\theta_R=33^{\circ}$,
$35^{\circ}$ and $37^{\circ}$. These angles correspond to directions
of the emission of the dibaryons with a certain mass.

Each telescope included two thin plastic scintillation $\Delta E$
detectors ($4\times 4\times 0.5$ cm$^3$)
and one $E$ detector of thick plastic ($5\times 5\times 20$ cm$^3$,
used in the right arm) or BGO crystal ($5\times 5\times 8$ cm$^3$, used
in the left arm). Each detector was viewed by a PMT-143 phototube coupled
to specially developed fast electronic modules.
A trigger was generated by 4-fold coincidence of the two
$\Delta E$ detector signals of the left arm
combined with those of any telescope of the right arm.
A time resolution better than 0.5 ns was achieved, which allowed
to suppress essentially an accidental coincidence background.
The scattered proton $E$ signals,
selected in the coincidence, formed the energy spectrum and accordingly the
missing mass spectrum. Each useful event including two time of flights
and two energies were stored event by event and then analyzed off line.

The elastic $pd$ scattering was measured
at various angles of the spectrometer arms. The proton energy
spectra obtained were used to
calibrate the spectrometer in the proton energy and, accordingly, in
the missing mass. The missing mass resolution of the spectrometer
was $\Delta M\approx 3$ MeV, the angular resolution was $1^{\circ}$.

\section{Results of the measurements \protect\\}
\label{sec:result}

\begin{figure}
\vspace*{5cm}
\epsfysize=16cm
\epsfxsize=16cm
\epsffile{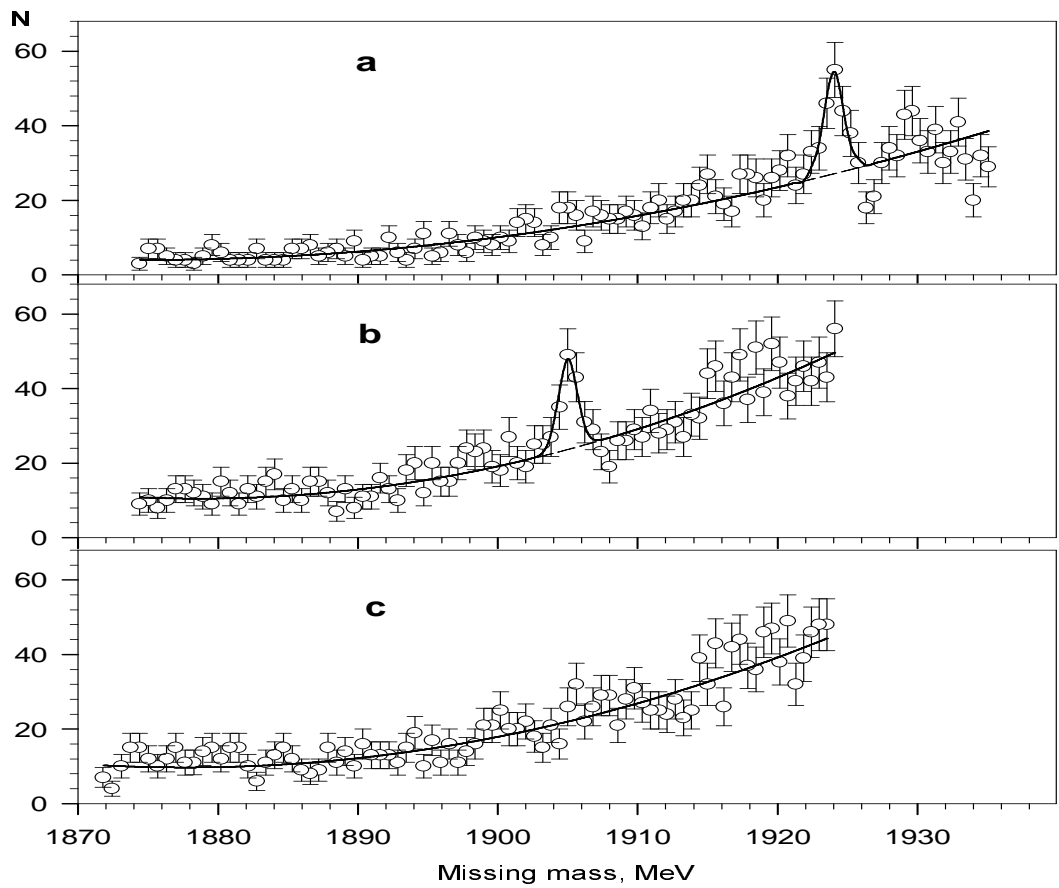}
\caption{The missing mass spectra of the reaction on CD$_2$:
(a) from the run data at
$\theta_L=70^{\circ}$ and $\theta_R=33^{\circ}$,
b) from the run data at $70^{\circ}$
and $35^{\circ}$, summarized with the run
data at $72.5^{\circ}$ and $33^{\circ}$,
c) from the run data at $70^{\circ}$
and $37^{\circ}$, summarized with the run data at
$72.5^{\circ}$ and $35^{\circ}$.
\label{fig3}}
\end{figure}

The experimental missing mass spectra obtained with the
deuteried polyethylene target are shown in Figs.~\ref{fig3}(a,b,c).
Each spectrum corresponds to a certain combination of outgoing angles
of the scattered proton and the second charged particle.
These combinations in Figs.~\ref{fig3}(b,c) are consistent with the shift
of the emission angle $\theta_R$ of the dibaryon with the given mass
when the angle $\theta_L$ changes from $70^{\circ}$ to
$72,5^{\circ}$. As is evident from Figs.~\ref{fig3}(a) and
\ref{fig3}(b), a resonance-like behavior of the spectra
is observed for the CD$_2$ target
in two mass regions at $1905\pm 2$ MeV and $1924\pm 2$ MeV. On the other
hand, the experiment with the carbon target resulted in a rather
smooth spectra \cite{izv}.
This smoothness is caused by an essential increase of the contribution of
background reactions in the interaction of the proton with the carbon.

It is worth noting that the resonance structure at 1905 MeV
is observed in both runs at $\theta_L=70^{\circ}$ and
$\theta_L=72.5^{\circ}$ and at different setup modifications. This
essentially decreases the possibility of a random origin of the
observed structure. The results obtained at different $\theta_L$
were summed up to improve statistics.

The mass spectra obtained in Figs.~\ref{fig3}(a,b,c) were interpolated
by  second order polynomials (for the background) plus Gaussians
(for the peaks in Figs.~\ref{fig3}(a) and \ref{fig3}(b)). Let us regard
Fig.~\ref{fig3}(b). In this case the interpolation gave
$\chi^2/(n-5)=0.81$. We determine the number of standard deviations (SD)
as
\[
\frac{N_{eff}}{\sqrt{N_{eff}+N_{b}}},
\]
where $N_{eff}$ is the number of events above the background curve and
$N_{b}$ is the number of events below this curve.
Taking 5 points for the peak at $M=1905$ MeV, we have
\mbox{$65/\sqrt{185}=4.8$ SD}. That corresponds to the probability of
statistic fluctuations $P$ equal to $2.7\times 10^{-5}$ \cite{stat}.

Using the same procedure for the peak at $M=1924$ MeV in Fig.~\ref{fig3}(a)
results in 4.9 SD and $P=1.6\times 10^{-5}$.

If, for the number of SD, we use  the expression \cite{tat}
\[
\frac{\sum_{i}^{n=5} (N_{ti}-N_{bi})/\sigma_i^2}
{\sqrt{\sum_i^{n=5} 1/\sigma_i^2}}
\]
where $N_{ti}$, $N_{bi}$, and $\sigma_i$ correspond, respectively, to total,
background, and error bar data,
then we obtain 4.2 S.D. for the peak at $M=1905$ MeV and 4.4 S.D. for
the peak at $M=1924$ MeV.

The widths of both observed peaks correspond to the experimental
resolution (3 MeV).

The peak at 1924 MeV was obtained only for one spectrum close
to the upper limit of the missing mass. In the other cases this mass
position was beyond the range of measurement.
Therefore, in the present work we analyze in more detail
the peak at 1905 MeV only.

The experimental missing mass spectra in the range of
1872--1914 MeV, after
subtracting the carbon contributions, are shown in Figs.~\ref{fig4}(a,b,c).

As seen from Figs.~\ref{fig3} and \ref{fig4}, the resonance behaviour of
the cross section exhibits itself in a limited angular region.

\section{Analysis of the experimental data \protect\\}
\label{sec:analys}

If the observed structure at $M=1905$ MeV corresponds to a
dibaryon decaying mainly into two nucleons, then the expected
angular cone size of emitted nucleons would be about 50$^{\circ}$.
Moreover, the angular distributions of the emitted
nucleons are expected to be very smooth in the angular region under
consideration. Thus, even assuming that the
dibaryon production cross section is equal to that of elastic scattering
 (40 $\mu b$/sr) its contribution to the missing mass spectra in
Figs.~\ref{fig3}(a,b,c) would be nearly the same and would not
exceed 1--2 events. Hence, the peaks found are hardly interpreted as
a manifestation of the formation and decay of such states.

It was shown in \cite{fil2} that the decay of the SND into
$\gamma NN$ had to be characterized in the rest frame
by a narrow peak near the maximum photon energy in
the probability distribution of the
dibaryon decay over an emitted photon energy. It leads to an essential
limitation of the
outgoing nucleon angles. On the other hand, if such a dibaryon
decays into $\gamma d$, the emitted deuteron angles are limited
by the following condition: $\sin\theta_d\le M p^*_d/(m_d p_D)$,
where $p_D$ is the momentum (in the lab. syst.)
of the dibaryon and $p^*_d$ is the momentum
(in the c.m.s.) of the deuteron.

Using the Monte Carlo simulation we estimated the contribution
of the SNDs with different quantum numbers and
$M$=1905 MeV to the mass spectra at various angles of the
left and right arms of our setup.
The production cross section and branching ratio of these
states were taken according to the calculations presented in
Sec.\ \ref{sec:2} and the proton beam current was assumed to be
equal to 0.1 nA.
The results obtained for different quantum numbers and decay modes
of the dibaryons are listed in Table \ref{table2}.

This calculation showed that the angular cone of protons and deuterons
emitted from a certain dibaryonic state can be rather narrow.
The axis of this cone is lined up with the direction of the dibaryon
emission. Therefore, placing the right spectrometer arm at the
expected angle of the dibaryon emission we increase essentially
the signal-to-background ratio.

Figs.~\ref{fig5}(a,b) exhibit the angular distributions of charged
particles (either $p$ or $d$) from the decay of the dibaryon $D(1,1^+)$
(Fig.~\ref{fig5}(a)) and $D(0,0^+)$ (Fig.~\ref{fig5}(b)).
The solid and dashed curves correspond to the
emission angles of the scattered proton equal to $70^{\circ}$ and
$72.5^{\circ}$, respectively. The solid vertical lines show the location
of the right arm detectors. These figures demonstrate the dependence of
the angular distribution of the charged particles under consideration
on the quantum numbers of the dibaryons.

\begin{figure}
\vspace*{5cm}
\epsfysize=16cm
\epsfxsize=16cm
\epsffile{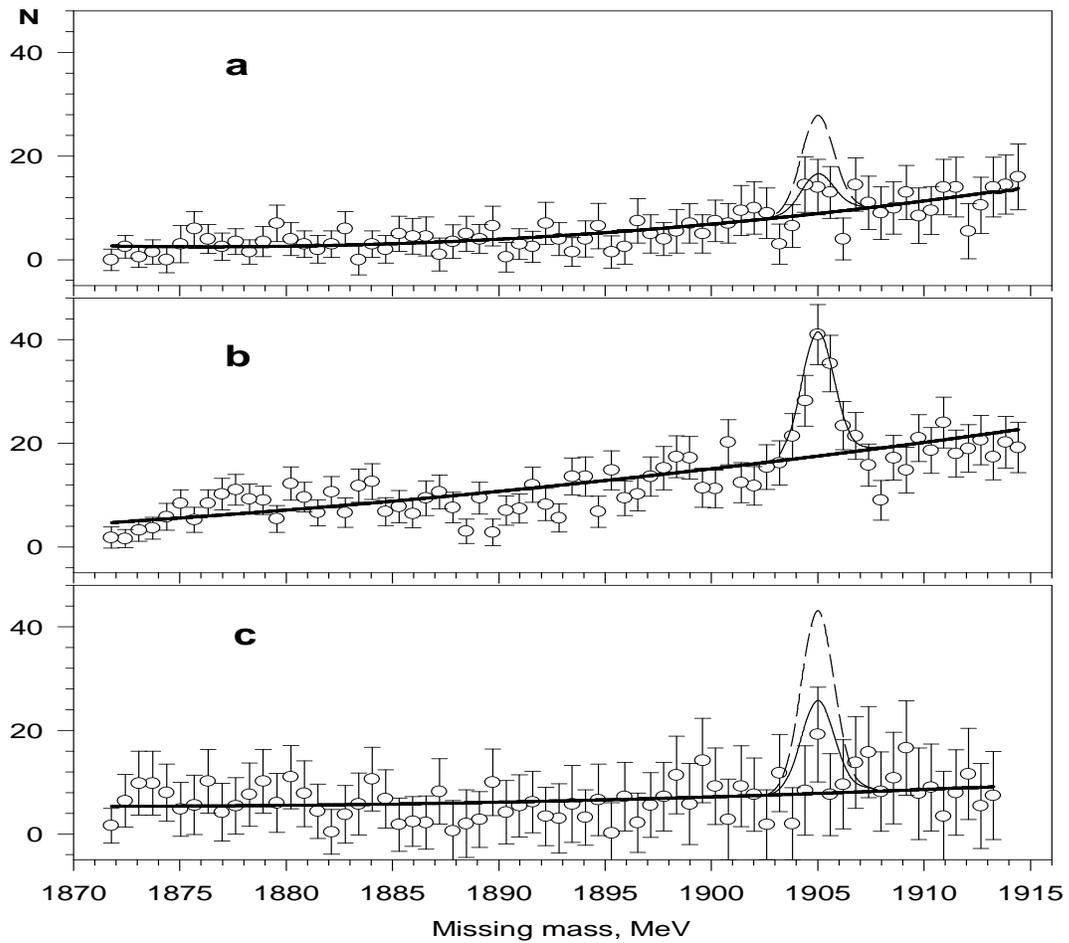}
\caption{The missing mass spectra of the reaction on deuteron.
Areas marked by letters a), b), c) correspond to the same
experimental conditions as in Fig.3. The solid and dashed curves
are normalized theoretical predictions of the yields of the supernarrow
dibaryon with isotopic spin equal to 1 and 0, respectively.}
\label{fig4}
\end{figure}

\begin{figure}
\epsfysize=16cm
\epsfxsize=15cm
\epsffile{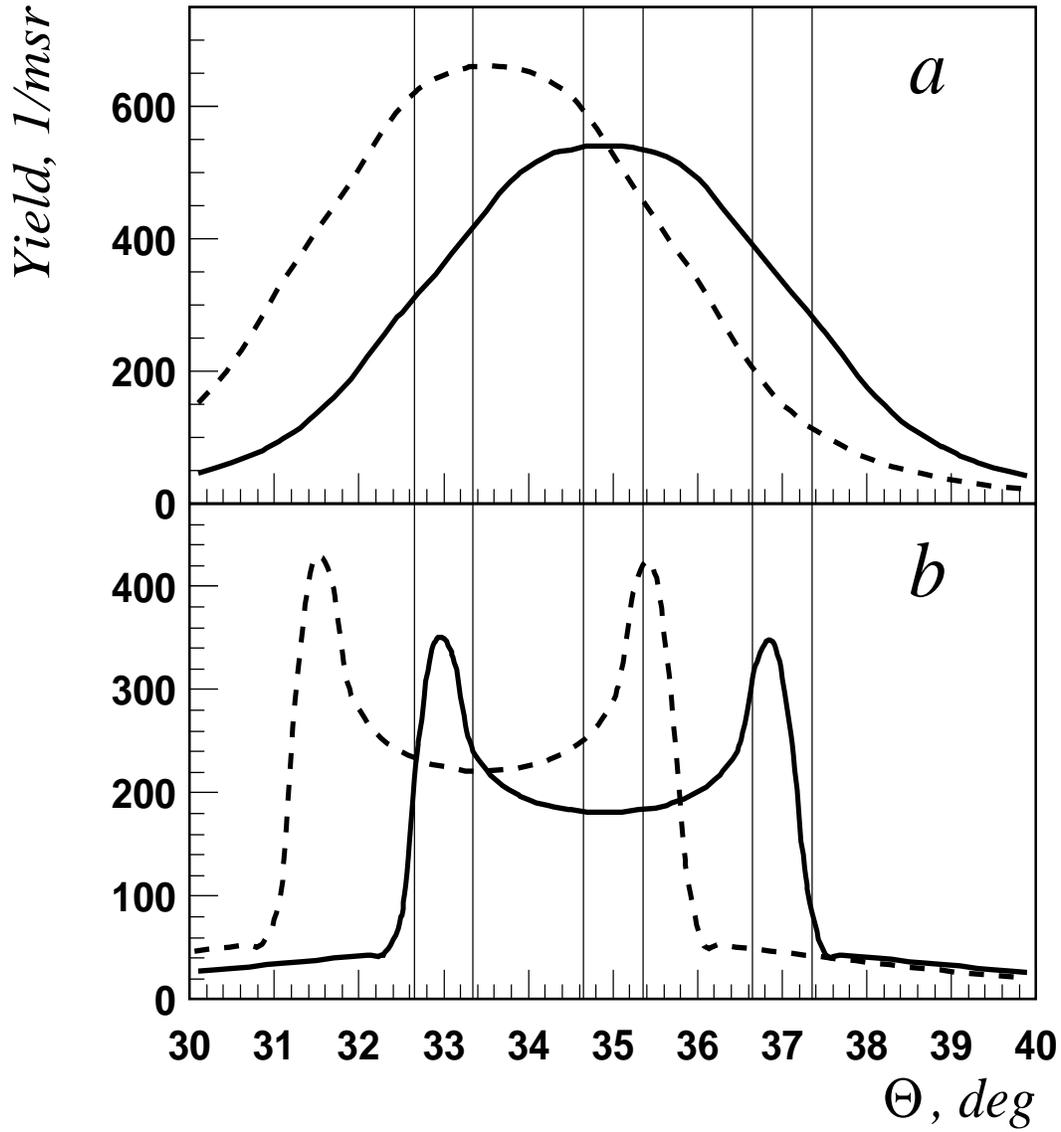}
\caption{ The angular distribution of protons emitted by the dibaryon
$D(1,1^+)$ (a), and the sum of the angular distributions of protons
and deuterons emitted by the dibaryon $D(0,0^+)$ (b).
The solid and dashed curves correspond to
the scattered proton angles $70^{\circ}$ and $72.5^{\circ}$, respectively.
The solid vertical lines show the location of the right arm detectors.}
\label{fig5}
\end{figure}

In Figs.~\ref{fig4}(a,b,c), the experimental spectra are compared with
the predicted yields normalized to the maximum of the measured signal in
Fig.~\ref{fig4}(b). The solid and dashed curves in these figures correspond
to the states with the isospin $T$=1 and $T$=0, respectively.
The background is described by a second order polynomial.

As is seen from these figures and Table \ref{table2}, the ratios
of the calculated yields to the given spectra are expected to be
$0.3:1: 0.7$, if  the state at 1905 MeV is interpreted as an isovector
dibaryon ($D(T=1,J^P=1^{+})$ or $D(1,1^-)$).
This is in agreement with our experimental data within the
errors. On the other hand, the signals from isoscalar
dibaryons ($D(0,0^+)$ or $D(0,0^-)$) could be
observed in Figs.~\ref{fig4}(b,c) with the same probability.
So, the dibaryon with $M=1905$ MeV is most likely to have the isospin
equal to 1.

In order to estimate the dibaryon production cross section,
we use normalization to the elastic $pd$ scattering and then,
comparing the theoretical predictions for the yields
of one of the isovector dibaryons with the experimental data,
find that the differential cross section of this dibaryon production
is equal to $8\pm 4\;\mu b/sr$. That allows us to evaluate the product of
unknown constants in Eqs.\ (\ref{one})
and (\ref{two}). If it is a dibaryon $D(1,1^+)$, then
$\eta g_1^2/4\pi=(0.44\pm 0.22)\times 10^{-2}$. But if it is a
dibaryon $D(1,1^-)$, we have $\eta g_2^2/4\pi=(0.2\pm 0.1)\times 10^{-2}$.
The quantum numbers $J^P$ of the observed state could be found,
for example, by a search for such dibaryons in the processes
of charged pion photoproduction by polarized photons from the deuteron
\cite{alek1}.

As for the structure at 1924 MeV, the number of events in this peak
exceeds the expected yields of the usual dibaryons
(decaying into two nucleons) by almost two orders, even when produced
with the rather big cross
section of 40$\mu b/sr$. So this state, probably, could be a
supernarrow dibaryon, too.

It should be noted that the reaction $pd\to NX$ was investigated in
other works, too (see, for example, \cite{set}). However, in contrast to
the present work, the authors of these works did not study either
the correlation between the parameters of the scattered nucleon and
the second
detected particle or the emission of the photon from the dibaryon decay.
Therefore, in these works the relative contribution of the dibaryons under
consideration was small, which hampered their observation.

\section{Conclusions \protect\\}
\label{sec:concl}

The following conclusions can be made .

1) As a result of the study of the reactions $pd\to ppX_1$ and $pd\to pdX_2$
two narrow peaks at 1905 and 1924 MeV with  widths
less than 3 MeV have been observed in the missing mass spectra.

2) The analysis of the angular distributions of the protons and the
deuterons from decay of particle $X$, produced in the reaction
$pd\to pX$, showed that the peak found at 1905 MeV can be
explained as a manifestation of the SND
with isospin equal to 1, the decay of which into two nucleons
is forbidden by the Pauli exclusion principle.

3) Probably, the observed state at 1924 MeV can be interpreted as
the SND, too.

\acknowledgments

We thank T.E. Grigorieva and Yu.M. Burmistrov for their active participation
in the creation of the setup and in the experimental runs.

\newpage
\begin{table}
\caption{ Decay widths of the dibaryons $D(0,0^+)$, $D(0,0^-)$,
$D(1,1^+)$ and $D(1,1^-)$
at various dibaryon masses $M$.
\label{table1}}
\begin{tabular}{dddddddcc}
   & \multicolumn{3}{c}{$D(0,0^+)$}&\multicolumn{3}{c}{$D(0,0^-)$}&
\multicolumn{1}{c}{$D(1,1^+)$}&\multicolumn{1}{c}{$D(1,1^-)$} \\
\tableline 
\multicolumn{1}{c}{$M$}&\multicolumn{1}{c}{$\a_{\g d}$}&
\multicolumn{1}{c}{$\a_{\g pn}$}&\multicolumn{1}{c}{$\a_t$}&
\multicolumn{1}{c}{$\a_{\g d}$}&\multicolumn{1}{c}{$\a_{\g pn}$}&
\multicolumn{1}{c}{$\a_t$}&\multicolumn{1}{c}{$\a_t\approx\a_{\g NN}$}&
\multicolumn{1}{c}{$\a_t\approx\a_{\g NN}$} \\
\multicolumn{1}{c}{(GeV)}&\multicolumn{1}{c}{(keV)}&
\multicolumn{1}{c}{(keV)}&\multicolumn{1}{c}{(keV)}&
\multicolumn{1}{c}{(keV)}&\multicolumn{1}{c}{(keV)}&
\multicolumn{1}{c}{(keV)}&\multicolumn{1}{c}{(eV)}&
\multicolumn{1}{c}{(eV)} \\ \tableline
1.90& 0.005& 0.010& 0.015& 0.001 & 0.002 & 0.003&  0.51& 0.13 \\
1.91& 0.01 & 0.05 & 0.06 & 0.003 & 0.007 & 0.010& 1.57&0.39 \\
1.93& 0.05 & 0.18 & 0.23 & 0.012 & 0.033 & 0.045&  6.7& 1.67 \\
1.96& 0.17 & 0.80 & 0.97 & 0.04  & 0.14  & 0.18 &  25.6& 6.4  \\
1.98& 0.3  & 1.6  & 1.9  & 0.08  & 0.27  & 0.35 &  48 & 12    \\
2.00& 0.5  & 2.7  & 3.2  & 0.13  & 0.47  & 0.60 &  81 & 20  \\
2.013&0.7  & 3.6  & 4.3  & 0.17  & 0.64  & 0.81 &  109 & 27 \\
\end{tabular}
\end{table}

\begin{table}
\caption{The expected contributions of the SNDs
with $M$=1905 MeV and different quantum numbers to
the spectra of mass at various angles of the left and right arms of
the setup.
\label{table2}}
\begin{tabular}{llrrrrrrrrrr}
\multicolumn{2}{r}{$\theta_L$}&\multicolumn{5}{c}{$70^{\circ}$}&
\multicolumn{5}{c}{$72.5^{\circ}$} \\ \tableline
\multicolumn{2}{r}{$\theta_R$}&$31^{\circ}$&$33^{\circ}$&$35^{\circ}$
&$37^{\circ}$&$39^{\circ}$&$31^{\circ}$&$33^{\circ}$&$35^{\circ}$
&$37^{\circ}$&$39^{\circ}$ \\
\multicolumn{1}{l}{$T,J^P$}&\multicolumn{11}{c}{} \\ \tableline
$0,0^+$& $\gamma pn$ & 5&  8&  8&  7&  5&  9&  10&  10&  6& 3 \\
$0,0^+$& $\gamma d $ & 0& 38& 17& 32&  0&  4&  23&  34&  0& 0 \\
$0,0^-$& $\gamma pn$ & 3&  4&  5&  4&  2&  4&   6&   5&  2& 2 \\
$0,0^-$& $\gamma d $ & 0& 19& 10& 17&  0&  2&  12&  17&  0& 0 \\
$1,1^+$& $\gamma pn$ &14& 55& 81& 50& 12& 47&  97&  79& 23& 5 \\
$1,1^-$& $\gamma pn$ &30&120&180&110& 27&103& 213& 174& 50&12 \\
\end{tabular}
\end{table}

\newpage

\newpage

\newpage

\newpage

\newpage

\end{document}